# Evidence for a Preferred Handedness of Spiral Galaxies

Michael J. Longo[1]
Department of Physics, University of Michigan, Ann Arbor, MI 48109

In this article I extend an earlier study of spiral galaxies in the Sloan Digital Sky Survey (SDSS) to investigate whether the universe has an overall handedness. A preference for spiral galaxies in one sector of the sky to be left-handed or right-handed spirals would indicate a parity-violating asymmetry in the overall universe and a preferred axis. The previous study used 2616 spiral galaxies with redshifts <0.04 and identified handedness. The new study uses 15158 with redshifts <0.085 and obtains very similar results to the first with a signal exceeding 5 $\sigma$, corresponding to a probability ~$2.5 \times 10^{-7}$ for occurring by chance. A similar asymmetry is seen in the Southern Galaxy spin catalog of Iye and Sugai. The axis of the dipole asymmetry lies at approx. ($l, b$) =(52°, 68.5°), roughly along that of our Galaxy and close to alignments observed in the WMAP cosmic microwave background distributions.

## 1. INTRODUCTION

Symmetry has a strong appeal to the human psyche. Nature, however, exhibits some surprising asymmetries. On the smallest scales, an asymmetry (parity violation) was found in the angular distribution of electrons in the beta decay of spin-oriented $^{60}$Co, confirming the proposal by Lee and Yang (1956) that parity was violated in weak decays. On the molecular scale, there is a large predominance of left-handed amino acids over right-handed ones in organisms, the origin of which is still not well understood. It is reasonable to ask if nature exhibits such an asymmetry on the largest scales.

Spiral galaxies with a well-defined handedness offer a means to test this possibility. Ideally the signal for such an asymmetry would be an excess of one handedness or "spin" in a large region of the sky and a similar excess of the other handedness in the opposite direction (i.e., a dipole). The preponderance of data in the northern Galactic hemisphere, as well as the masking of much of the sky by dust in the Milky Way, complicates the search for such an effect. However, the spiral handedness technique has important advantages in that it is not biased by the incompleteness of the maps or by atmospheric or instrumental effects, which cannot turn right-

[1] email: mlongo@umich.edu

handed spirals into left-handed. One has to be careful of an overall bias due to a preference toward assigning left-handed or right-handed. Such a bias would show up as a "monopole". In principle, a computer algorithm to assign handednesses could be developed (Longo 2007a, b), but in practice the huge range of galaxy brightness, color, size on the sky, orientation, and structure makes this exceedingly difficult to do efficiently[2]. In this study, as in the Galaxy Zoo study (Land et al. 2008), human scanners chose the spiral sample and made the handedness assignments. Precautions against such a left/right bias will be discussed below.

In the first study (Longo 2007a, b) I used galaxies from the SDSS DR5 database (Adelman-McCarthy et al. 2007) that contains ~40,000 galaxies with spectra for redshifts $z < 0.04$. In this study I use the DR6 database (Adelman-McCarthy et al. 2008) with ~230,000 galaxies with $z < 0.085$. A few percent of these are spiral galaxies with identifiable handedness that can be used in the study.

## 2. THE ANALYSIS

Objects classified as "galaxies" in the SDSS DR6 database were used in this analysis. A list of galaxies that had measured redshifts less than 0.10 was obtained from the SDSS DR6 web sites, cas.sdss.org/dr6 and casjobs.sdss.org. Spiral galaxies are typically bluer than elliptical ones. Strateva et al. (2001) show that elliptical galaxies can be separated from spirals fairly cleanly by a $z$-dependent cut on ($U$-$R$) where $U$ and $R$ are the apparent magnitudes for the ultraviolet (354 nm) and red (628 nm) bands respectively (York et al. 2000). A conservative cut to enhance the fraction of spirals was therefore made by requiring that ($U$-$R$) < 2.85. Galaxy images from the resulting list of ~200,000 galaxies were then looked at by a team of 5 scanners.

Individual RGB images of the galaxies from the list were acquired from the SDSS web site and displayed to the scanners using an HTML/JAVA program. The HTML program mirrored half of the images at random to reduce scanning biases favoring a particular handedness.[3] The scanners had no visual cue as to whether the image was mirrored. Scanners were assigned small $z$ slices at random, and the scanning was done in random order with respect to right ascension, $\alpha$, and declination $\delta$ so that any scanning bias could not cause a systematic bias in the ($\alpha$, $\delta$, $z$) distributions of the handedness, only a possible overall bias in the complete sample. The scanners had only 3 choices: *Left*, *Right*, or *Unclear*, where *Left* ≡ ↺ and *Right* ≡ ↻. No attempt to oth-erwise classify the galaxies was made. Scanners were instructed to classify galaxies as Unclear unless the handedness was clear. Overall, about 15% of the galaxies were classified as having recognizable handedness (*L* or *R*). No further analysis of the *U*'s was done.

---

[2] A program using a rotating mask is described in Longo 2007a. It gave similar results to the human scanning used in later versions, but was considerably less efficient, especially at larger redshifts.

[3] Ideally, to eliminate biases completely, the complete lists should be scanned twice, intermingling all the images with their mirrored versions, but this would have taken another year of work. No record of the mirroring was kept.

To reproduce the earlier study (Longo 2007a, 2007b) as closely as possible, I required that the green magnitude be <17 for redshifts $z < 0.04$. Beyond $z=0.04$ the magnitude limit was increased to 17.4. The handedness of galaxies fainter than magnitude 17.4 and galaxies with $z > 0.085$ were generally difficult to classify, so these were not used.

As in the previous studies, a cut was made to remove the bluest galaxies that tend to be those with recent star formation initiated by a collision. This required (*U-Z*) > 1.6, where *Z* is the apparent magnitude in the far infrared; it removed 1.9% of the *L+R* sample. A cut to remove the reddest galaxies, (*U-Z*) < 3.5 was also made; this removed 2.7% of the sample, leaving 15158.

Most of the SDSS DR6 data are in the northern Galactic hemisphere ($\alpha \sim 192°$, $\delta \sim 27°$) with declinations -5°< $\delta$ <63°. In the southern Galactic hemisphere there is only coverage in 3 narrow bands in $\delta$ near $\delta=-10°, 0°$, and 14°, each about 4° wide. In the earlier analysis I used all declinations between -19° and +60° which includes most of the data. In this analysis I use the same declination ranges. This left 14683 galaxies compared to 2616 in the first study.

None of these cuts or the incomplete coverage of the survey would be expected to cause a bias between left- and right-handed spirals.

## 3. RESULTS

A plot of asymmetries $\langle A \rangle \equiv (R-L)/(R+L)$, binned in 30° sectors of right ascension and 0.01 slices in *z* for *z*<0.085, is shown in Fig. 1. Positive $\langle A \rangle$ are shown in red and negative ones in blue. The larger numbers near the perimeter give the net asymmetry for the entire right ascension sector. The black numbers in parentheses next to them give the total number of galaxies in that sector. The $\sigma$ are determined from standard normal distribution statistics, $\sigma(N) = \sqrt{N}$, which gives $\sigma(\langle A \rangle) = 1/\sqrt{R+L}$. There is an apparent excess of left-handed spirals in the sectors for 150°< $\alpha$ <240° and a complementary excess of right-handed in the opposite hemisphere, though there are only 1/7[th] as many galaxies there.

The incompleteness of the survey, especially in $\delta$, makes a complete multipole analysis of the asymmetry data of dubious value. In any case a preferred spiral handedness implies a dipole component and the lack of a monopole (bias), so I restrict my analysis to these two terms.

*Bias*–The galaxies were scanned in random order with respect to ($\alpha$, $\delta$, $z$), so no ($\alpha$, $\delta$) dependent bias is possible. Also, half the galaxies were randomly mirrored during scanning and precautions against left-right bias were taken in the web interface used by the scanners. The best check for an overall bias is to look at the complete scanned sample that included galaxies at larger *z* and fainter luminosities as well as some that were scanned more than once. This sample included 25612 galaxies and gave $R$=12707, $L$=12905 and an overall asymmetry of $-0.0077 \pm 0.0062$, even though this sample included 7 times as many galaxies in the $150° < \alpha < 240°$ sector with its apparent excess of $L$ spirals. If the $150° < \alpha < 240°$ sector is removed, the asymmetry becomes $+0.0133 \pm 0.010$, consistent with no bias or a small positive one.

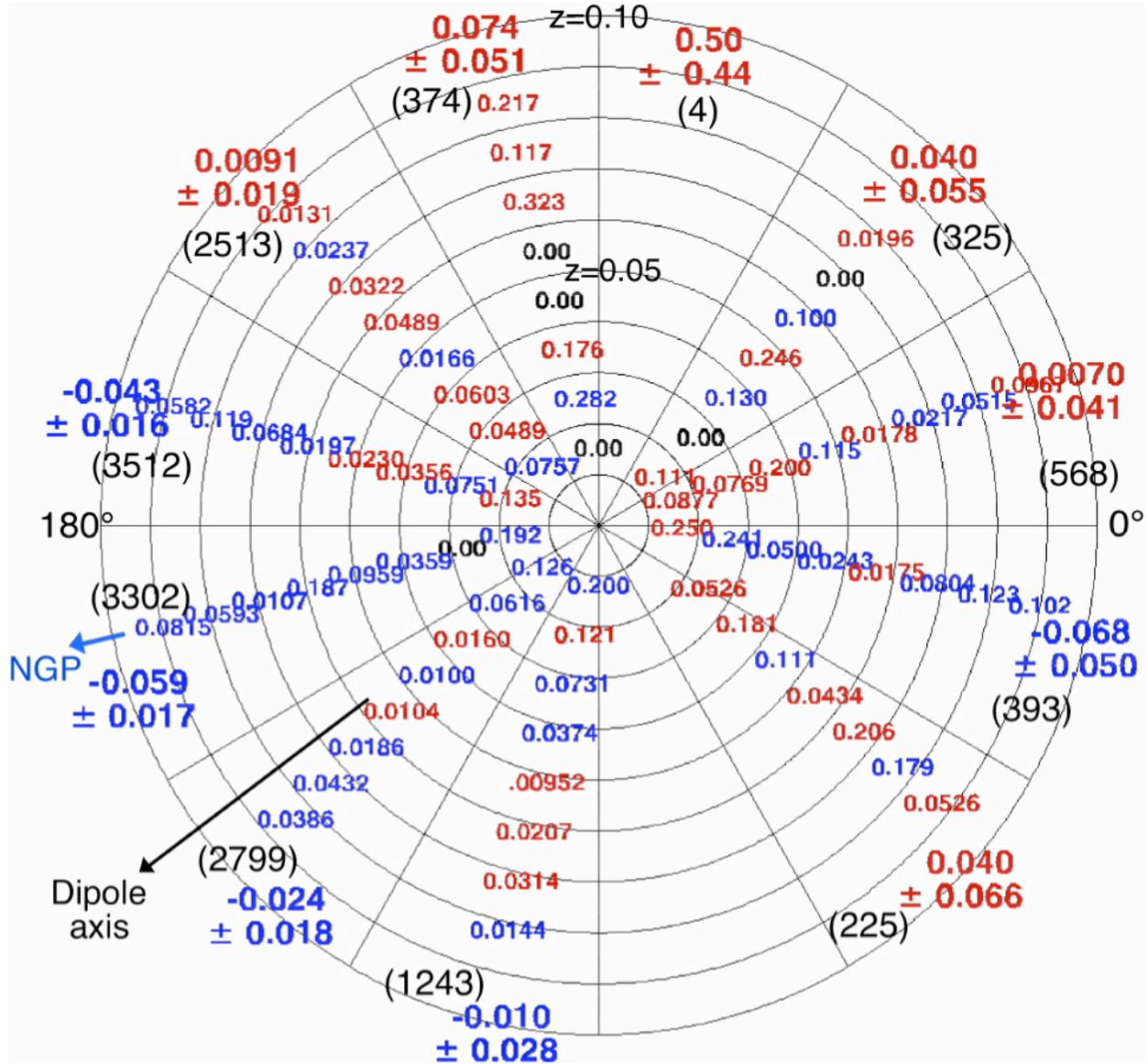

FIG. 1. Polar plot of net asymmetries ⟨A⟩ in 30° sectors in right ascension and slices in $z$. Segments with positive ⟨A⟩ are indicated in red and negative ⟨A⟩ in blue. The ⟨A⟩ for segments with <10 galaxies are not shown. The larger numbers near the periphery give the overall asymmetry for that sector; the black numbers in parentheses are the total number of spiral galaxies in the sector. The NGP is the north pole of our Galaxy, so that the left half of the plot corresponds roughly to the northern Galactic hemisphere. The black arrow shows the most probable dipole axis. Declinations between -19° and +60° were used.

*Asymmetry*–In the earlier analysis, in order to determine the overall statistical significance of the apparent asymmetry I used data in the declination range $-19° < \delta < 60°$, and the right ascension ranges $195° \pm 45°$ and $0° \pm 40°$. Here I use the same $\delta$ range and right ascensions in a narrower range $195° \pm 30°$ and $15° \pm 30°$. Table I shows the resulting asymmetries and their uncertainties. The $165° < \alpha < 225°$ sector with 86% of the galaxies shows an asymmetry of $-0.0695 \pm 0.0127$, a $5.48\sigma$ effect. The data in the sparsely covered southern Galactic hemisphere ($-15° < \alpha < 45°$) show a small positive asymmetry as would be expected for a real signal. Overall the asymmetry is $-0.0607 \pm 0.0118$, a $5.15\sigma$ effect with a probability of $2.5 \times 10^{-7}$ for occurring by chance. The completely new data alone with $0.04 < z < 0.085$ gives an asymmetry $-0.0569 \pm 0.0142$ with a probability of $6.0 \times 10^{-5}$. For $z<0.04$ the asymmetry is $-0.0692 \pm 0.0212$ with a probability of $1.1 \times 10^{-3}$. Thus there is no indication of a $z$ dependence of the asymmetry.

Table I. Number counts and net asymmetries $\langle A \rangle = (N_R - N_L)/N_{Tot}$ for the right ascension ranges indicated. The last two columns give the number of standard deviations for the $\langle A \rangle$ and the probability.

| $\alpha$ Range | $N_R$ | $N_L$ | $N_{Tot}$ | $\langle A \rangle \pm \sigma$ | $\langle A \rangle / \sigma$ | Prob. |
|---|---|---|---|---|---|---|
| -15° to 45° | 495 | 490 | 985 | 0.005±0.032 | +0.16 | 0.87 |
| 165° to 225° | 2890 | 3322 | 6212 | −0.0695±0.0127 | −5.48 | $2.1 \times 10^{-8}$ |
| Overall | | | 7197 | -0.0607±0.0118 | −5.15 | $2.5 \times 10^{-7}$ |

*The Dipole*–A real large-scale spiral asymmetry would exhibit itself as a dipole with a $\cos\gamma$ dependence where $\gamma$ is the space angle between the position of the galaxy and the axis of the dipole. This $\cos\gamma$ behavior is not affected by the incompleteness of the sample, but the sparseness of the data in one hemisphere makes fitting the $\cos\gamma$ dependence more difficult. To investigate a dipole, the complete sample of 15158 galaxies without position cuts was used. No *a priori* assumptions about the direction of the dipole axis or its magnitude were made. No binning of the data was used in the analysis, so that the results do not depend on the choice of bins. First a possible axis was chosen. The $\gamma_i$ of each of the galaxies were calculated for that axis and a handedness of +1 or -1 was assigned to each galaxy. The 15158 points were then fitted to an $a \cos\gamma_i$ dependence. The $(\alpha_A, \delta_A)$ of the axis was varied stepwise to find the axis that gave the minimum $\chi^2$/dof. The best fit was found at $(\alpha_A, \delta_A) = (217°, 32°)$, or $(l, b) = (52°, 68.5°)$ in Galactic coordinates.

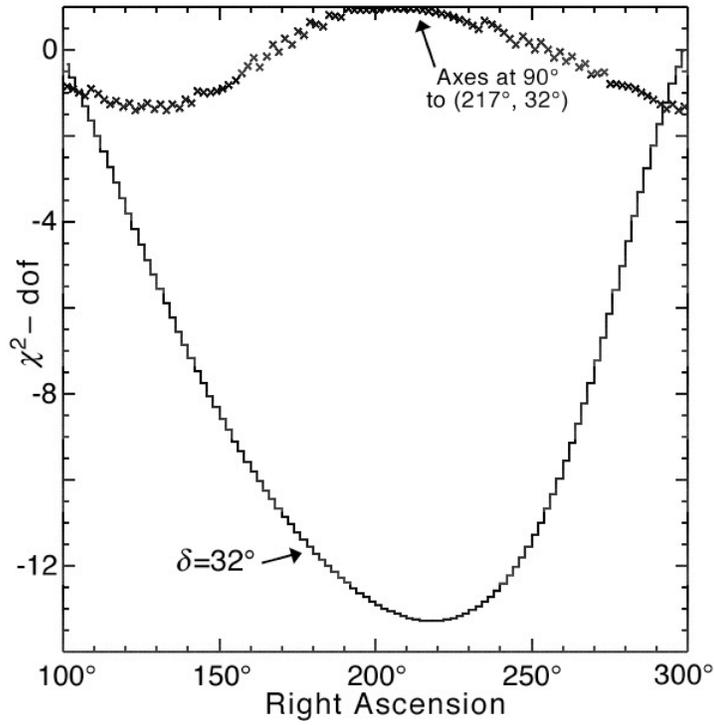

FIG. 2.–Variation of $\chi^2$–$dof$ with $\alpha_A$ for $\delta_A=32°$. The x's are for axes at 90° to the best-fit axis.

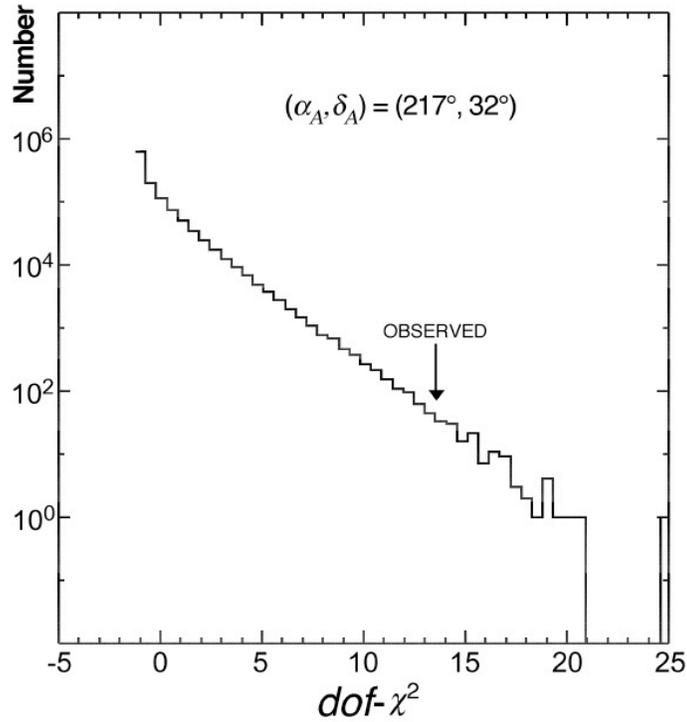

FIG. 3. – Probability of obtaining a particular value of ($dof$–$\chi^2$) for $1.2 \times 10^6$ samples of the 15158 galaxies with <u>randomized</u> handednesses and $(\alpha_A, \delta_A) = (217°, 32°)$. The arrow shows the value for the actual handedness assignments; the probability of finding that value or greater is $1.51 \times 10^{-4}$.

Fig. 2 shows the variation of ($\chi^2$–$dof$) with $\alpha_A$ for $\delta_A$=32°, the axis declination that gave the minimum $\chi^2$. The x's are the values obtained for axes that are in the plane orthogonal to the best-fit axis; these show a much smaller variation in $\chi^2$ and their average is close to 0. Note that while $\chi^2$ is a very well behaved function of $\alpha_A$ and $\delta_A$ its distribution is very unlike that of the usual $\chi^2$ distribution because the handedness can only take discrete values of +1 or –1 while the usual $\chi^2$ distribution is expected only for data that have a normal distribution. Thus in this case the statistical significance of the observed $\chi^2$ had to be determined numerically for the actual galaxy sample. The $\chi^2$ distribution that would be expected if the handednesses were distributed randomly was determined by generating 1.2 x $10^6$ samples of the 15158 galaxies with the handedness of each randomly assigned to either +1 or –1 with equal probability. The full range of $\alpha_A$ with $\delta_A$=32° was used. This gave the ($dof$–$\chi^2$) distribution shown in Fig. 3. The arrow shows the value for the actual handedness assignments; the probability of finding that value or greater is 1.51x$10^{-4}$.

The $a\cos\gamma_i$ fit gave $a$ = –0.0408 with an uncertainty $\sigma_a$ =±0.011. The randomly generated samples showed that $a/\sigma_a$ had a Gaussian distribution with an rms of $a/\sigma_a$=1.001 and a probability of $a/\sigma_a$ exceeding 0.0408/0.011 of 1.51x$10^{-4}$, the same as the probability from the $\chi^2$ distribution. The minimum in Fig. 2 is quite broad due to the sparse data in the southern Galactic hemisphere and the lack of data around $\gamma$~90° which corresponds to the Galactic plane. The uncertainty in space angle is estimated to be ~30° as determined from the angle at which the $\chi^2$ probability doubles from its minimum value.

## 4. COMPARISON WITH OTHER STUDIES

*Iye and Sugai* – Iye and Sugai (1991) have published a catalog of spin orientations of galaxies in the southern Galactic hemisphere that contains 8287 spiral galaxies. Of these, 3118 had $R$ or $L$ handedness about which both scanners agreed. I have analyzed their catalog using the sector –15°<$\alpha$<+45° and -60°< $\delta$ <+5°, directly opposite that used above[4]. Redshifts of most of their galaxies were not measured, so only their ($\alpha,\delta$) were used. This gave an asymmetry +0.047± 0.029 with a preponderance of right-handed spirals in the southern Galactic hemisphere, in excellent agreement with the asymmetry |$A$| = 0.0695±0.0127 that I observe for 165°<$\alpha$<225° with a preponderance of left-handed spirals. This provides an independent confirmation of a spin asymmetry at the 1.6$\sigma$ level. Without ($\alpha,\delta$) cuts their overall asymmetry was 0.000±0.014, consistent with no bias in their study. The combined probability for the dipole term found here and the asymmetry found for the Iye and Sugai catalog is 0.205(1.51x$10^{-4}$) ~ 3.1x$10^{-5}$.

---

[4] Note, however, that the Iye-Sugai catalog does not contain galaxies with declinations $\delta$ >-18°.

*Galaxy Zoo* – Galaxy Zoo (Lintott et al., 2008) is an online project in which >100,000 volunteers visually classify the morphologies of galaxies selected from the spectroscopic sample of the SDSS DR6, the same sample used here. In Land et al. 2008 they investigate the possibility of a large scale spin anisotropy. Each galaxy was classified an average of 39 times. Those galaxies for which over 80% of the votes agreed constituted their "clean" sample and over 95% their "superclean" sample. They found that there was a large *L/R* bias in their samples. Their clean sample contained 18471 *L* (clockwise) spirals and 17100 *R*; the superclean sample contained 7034 *L* and 6106 *R* spirals[5]. This gives an asymmetry (bias?) of $-0.0385 \pm 0.0053$ for the clean sample and a much larger bias of $-0.0706 \pm 0.0087$ for the superclean sample that presumably contained more clearly recognizable spirals. This should be compared to the upper limit of $-0.0077 \pm 0.0062$ for the bias found in this study as discussed in the section on biases above. Land et al. attributed these biases to the design of the Galaxy Zoo website or to a human pattern recognition effect that was shared by all their volunteers. A later bias study, mainly with the superclean sample, that compared monochrome images with mirrored RGB images found similar biases. They corrected for the bias by requiring only 78% agreement between scanners for the *R* galaxies in the clean sample and 94% in the superclean. They assumed the bias was independent of redshift and magnitude, despite the fact that it is much easier to correctly assign the handedness of larger, brighter galaxies. In their analysis they found a dipole term of about $2\sigma$ along $(\alpha, \delta)=(161°,11°)$ consistent with the axis I found in Longo 2007 at $(202°,25°)$. When a monopole (bias) term was also allowed, this became a $1\sigma$ effect.

It is difficult to compare this study with the Galaxy Zoo result. They used galaxies with redshifts up to 0.3, whereas I used those with $z<0.085$ and restricted the magnitude range because of the difficulty in assigning the handedness of fainter galaxies. Their large biases also caused large uncertainties in the monopole/bias term, while the biases in this analysis are consistent with 0.

## 5. DISCUSSION AND CONCLUSIONS

The galaxy handedness data show a strong signal for a preferred axis for redshifts <0.085. The new analysis reported here shows an asymmetry of $-0.0607 \pm 0.0118$, a 5.15 $\sigma$ effect with a probability of $2.5 \times 10^{-7}$ for occurring by chance. This analysis used data selection cuts very similar to those in the previous analysis and the sample contained about 7 times as many galaxies. The results of the two studies are in excellent agreement. The axis lies near $(\alpha_A, \delta_A) = (217°, 32°)$, or $(l, b) = (52°, 68.5°)$ in Galactic coordinates with the sense of the axis defined to be along the direction of the $L$ (↺) excess. The uncertainty in space angle of the axis is ~30°. An analysis for a dipole term made without prior assumptions or arbitrary binning showed a $\cos\gamma$ dependence of the asymmetries with a probability of occurring by chance of 1.5 x $10^{-4}$. The monopole (bias) term is consistent with 0. This result is consistent with the spin asymmetry of 0.045±0.035 with a probability of 20% found in an analysis of the Iye and Sugai spin catalog in the opposite hemisphere.

---

[5] See also Table 2 of Lintott et al. 2008  The uncertainties in the asymmetry are calculated as $\sigma(A) = 1/\sqrt{R+L}$.

This axis is 21.5° away from the north pole of our galaxy (NGP in Fig. 1) at $b$=90°. Our galaxy has its spin vector generally aligned with the preferred axis of spiral galaxies, corresponding to a probability for this to occur by chance of (1-cos 21.5°)/2 = 3.5%. There is no obvious redshift dependence of the asymmetry out to $z \approx 0.085$, which is well beyond the scale of superclustering[6]. Extension of this study to larger redshifts will be difficult due to the problem of reliably recognizing the handedness of faint galaxies.

There is now a vast literature on possible observations of cosmic anisotropies and explanations thereof. I shall only briefly comment on the observational effects. Many of these revolve around apparent anomalies in the Wilkinson Microwave Anisotropy Probe one-year, three-year (Hinshaw et al. 2007), and five-year data.[7] For example, analyzing the one-year data, Land and Magueijo (2005, 2007) find an unlikely alignment of the low $l$ multipoles and a correlation of azimuthal phases between $l$ = 3 and $l$ = 5 with an apparent axis ($l, b$) = (260°, 60°), an alignment they refer to as the "axis of evil". Using the three-year WMAP data, Copi et al. (2007) find a similar correlation of low $l$ multipoles and a significant lack of correlations for scales >60°. Analyzing the WMAP five-year and three-year data, Bernui (2008) finds a significant asymmetry in large-angle correlations between the north and south Galactic hemispheres at >90% confidence level, depending on the map used. Su and Chu (2009), in a recent reanalysis of the WMAP data, find a general alignment of the directions for $l$ = 2 to 10 modes to within about $1/4^{th}$ of the northern Galactic hemisphere at latitudes between about 45° and 85°. Similarly, Råth et al. (2009) find evidence for asymmetries and non-Gaussianities between the North and South Galactic polar regions. Campanelli, Cea, and Tedesco(2006, 2007) propose that the CMB quadrupole anisotropy is the result of an ellipsoidal universe with an axis at 50°< $l$ <125°, $b \approx$ 52°. It is notable that the axis I find at ($l, b$) =(52°, 68.5°) lies almost directly opposite the direction of the cold spot in the CMB found by Vielva et al. at ($l, b$) =(209°,-57°) with an angular size ~10°.

There is also a considerable literature on a "Virgo alignment" toward ($l, b$) ~ (281°, 75°) of axes related to the CMB as well as other anisotropic effects in polarizations of radio galaxies and the optical frequency polarization correlation of quasars. (See, for example, P. K. Samal et al. 2008 for a recent discussion of these effects.)

---

[6] For example, Virgo at the center of our local supercluster is at a redshift ~0.004.

[7] The WMAP data are available at http://lambda.gsfc.nasa.gov/product/map/current/m_products.cfm

All of these effects seem to lie at high Galactic latitudes. While the asymmetries in the CMB data are now well documented, there is always the possibility that they are due to a common systematic effect, in particular a bias from foreground contamination near the Galactic plane. Thus the fact that the spiral asymmetry axis also lies close to the Galactic poles provides significant evidence that the CMB asymmetries are real and suggests that both phenomena may have a common origin.

I am extremely grateful to the SDSS group whose efforts and dedication made this work possible. I am indebted to Dr. Masanori Iye for providing the spin catalog of southern galaxies. Undergraduates E. Mallen, A. Bomers, J. Middleton, and M. Pearce made important contributions in their diligent work as scanners. B. McCorkle did the JAVA/HTML programming.